\documentclass[aps,prx,amsfonts,amsmath,amssymb,raggedbottom,longbibliography,reprint,superscriptaddress,citeautoscript]{revtex4-2}

\usepackage{graphicx} 
\usepackage{float}
\usepackage{color}
\usepackage{bm}
\usepackage{hyperref}
\usepackage{todonotes}
\usepackage{verbatim}
\usepackage{soul}
\usepackage{glossaries}
\usepackage{sidecap}
\usepackage{hyperref}
\usepackage{ulem}
\hypersetup{
    colorlinks=true,
    linkcolor=blue,
    filecolor=magenta,
    urlcolor=blue,
    citecolor=blue,
}

\begin{document}

\title{Supermodulation-driven evolution of the nodal structure of bismuth-based cuprate superconductors}

\author{M. R. Norman}
\affiliation{Materials Science Division, Argonne National Laboratory, Lemont, IL 60439}

\date{\today}

\begin{abstract}
Recent work has shown novel properties of twisted cuprates.  In this paper, I point out that related phenomena occur intrinsically in bismuth-based cuprate superconductors due to the presence of the BiO supermodulation.  As the ratio of the supermodulation potential to the superconducting energy gap increases, two critical points are found where semi-Dirac nodes form (that is, that have quadratic dispersion in one direction and liner dispersion in the orthogonal direction).  The first critical point should be realized in Bi2212, the second in Bi2201.  Implications of these findings are discussed.
\end{abstract}

\maketitle

\section{Introduction}
Recently, it was predicted that unusual phenomena occur in cuprate superconductors if one layer is twisted relative to another \cite{Can21a,Can21b,Mercardo22,Song22,Volkov23a,Volkov23b,Volkov25} that has been extended to more than two twisted layers \cite{Tummuru22,Lucht25}.  This has in turn been studied in bismuth-based cuprates by two experimental groups \cite{Zhao23,Martini23}.  In this paper, I point out that related phenomena occur intrinsically in the bismuth-based cuprate superconductors due to the BiO supermodulation whose presence is felt in the CuO$_2$ planes. The essence behind the phenomena is that in both cases, twisting and supermodulation, the points at which the the Fermi surface crosses the twisted/translated Fermi surfaces (denoted as crossing points) have d-wave superconducting order parameters with opposite sign.  This opposing sign has been elucidated by photoemission experiments on Bi2212 (Bi$_2$Sr$_2$CaCu$_2$O$_{8+\delta}$) \cite{Gao24} and analyzed in subsequent work \cite{Zhan25}.

In this paper, I will explore the evolution of the superconducting electronic structure as the ratio of the supermodulation potential, $V$, with respect to the d-wave energy gap at the crossing points, $\Delta_c$, varies.  Two critical points are found.  As $V$ turns on, an energy minimum develops at the crossing point.  When $V \sim \Delta_c$ (denoted as $V_{c1}$), this minimum reaches zero energy and forms a semi-Dirac node, with quadratic dispersion normal to the Fermi surface and linear dispersion along the Fermi surface.  When $V$ increases beyond this, this semi-Dirac node splits into two nodes. One node ($L$) moves normal to the Fermi surface, the other ($T$) along the Fermi surface where it eventually merges with the d-wave node of one of the translated Fermi surfaces for $V \sim 2\Delta_c$ (denoted as $V_{c2}$).  This is again a semi-Dirac node, but this time with quadratic dispersion along the Fermi surface and linear dispersion normal to the Fermi surface.  For larger $V$ ($V > V_{c2}$), this semi-Dirac node is lifted and one reverts to a situation where the number of nodes in the Brilouin zone is equal to that for $V < V_{c1}$.  Consequences of this for both Bi2201 (Bi$_2$Sr$_2$CuO$_{6+\delta}$) and Bi2212 will be discussed at the end.

\section{Formalism}

\begin{figure}
\includegraphics[width=\columnwidth]{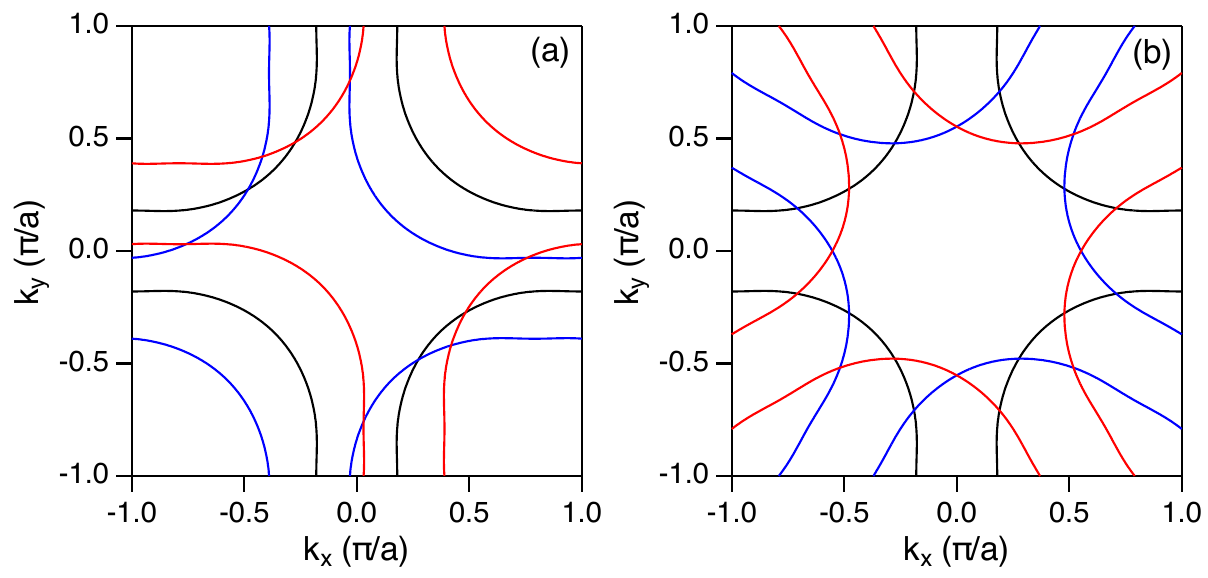}
\caption{Fermi surfaces for (a) the supermodulation case and (b) the trilayer twisted case.  Black is the original surface, blue and red are the translated/twisted surfaces. For (a), the supermodulation vector is $Q=(0.21,0.21)$ and for (b), the twist angle is $\phi=30^\circ$.  In these plots, $V=0$.}
\label{fig1}
\end{figure}

In the bismuth-based cuprates, there is a supermodulation wavevector due to the mismatch of the lattice constants of the BiO and CuO$_2$ layers which causes the BiO layers to buckle.  This wavevector in tetragonal notation is $Q=(0.21,0.21)\pi/a$ which is equivalent to $(0,0.21)\pi/a_O$ in orthorhombic notation where $a$ is the tetragonal lattice constant (Cu-Cu spacing) and $a_O \sim \sqrt{2}a$.  The resulting Fermi surface (black) and its translated $\pm Q$ variants (blue and red, respectively) are plotted in Fig.~\ref{fig1}a and contrasted with a related twisted case in Fig.~\ref{fig1}b.  For the latter, a trilayer case was considered where the middle layer stays fixed, the top layer is rotated by $\phi$, and the bottom layer rotated by $-\phi$, where $\phi = 30^\circ$ was chosen to get crossing points that approximately match the supermodulation case (with the blue and red surfaces being rotated by $\pm \phi$, respectively).  In these plots, a tight binding dispersion is used based on photoemission data \cite{phen95}. The effect of the supermodulation potential $V$ (interlayer potential in the twisted case) is to create avoided crossings at these crossing points.  When comparing these two plots, note two differences.  In the supermodulation case, the crossings only happen in the $X$ quadrants of the zone (upper left and lower right quadrants in Fig.~\ref{fig1}a).  In the twisted case, they occur in all quadrants, and also there are more crossings, so the net result in that the twisted case has four times the number of crossings in the entire zone.  The other important point is that upon inversion, the color of the Fermi surface inverts in the supermodulation case (blue and red interchange), but the color is preserved in the twisted case.

In the superconducting state, the secular equation for the Bogoliubov dispersions is:
\begin{equation}
\begin{split}
\hat{\mathcal{H}}(\bm{k})=
 &\left( 
 {\begin{array}{cccccc}
 \epsilon_k & V & V & \Delta_k & 0 & 0 \\
 V & \epsilon_{k+Q} & 0 & 0 & \Delta_{k+Q} & 0 \\
 V & 0 & \epsilon_{k-Q} & 0 & 0 & \Delta_{k-Q} \\
 \Delta_k & 0 & 0 & -\epsilon_k & -V & -V \\
 0 & \Delta_{k+Q} & 0 & -V & -\epsilon_{k+Q} & 0 \\
 0 & 0 & \Delta_{k-Q} & -V & 0 & -\epsilon_{k-Q}
 \end{array}} \right)
\end{split}
 \label{eq1}
\end{equation}
For the twisted case, one simply replaces ${k\pm Q}$ by $\hat{R}(\pm \phi)k$ where $\hat{R}$ is the in-plane rotation operator.  The important point is at the Fermi surface crossing points (where a black surface crosses a blue or red surface in Fig.~\ref{fig1}), the d-wave order parameters are equal but with opposite sign in both the supermodulation and twisted cases.  From here on, we will just treat the supermodulation case, as this is the focus of the paper.

\section{Results}

\begin{figure}
\includegraphics[width=\columnwidth]{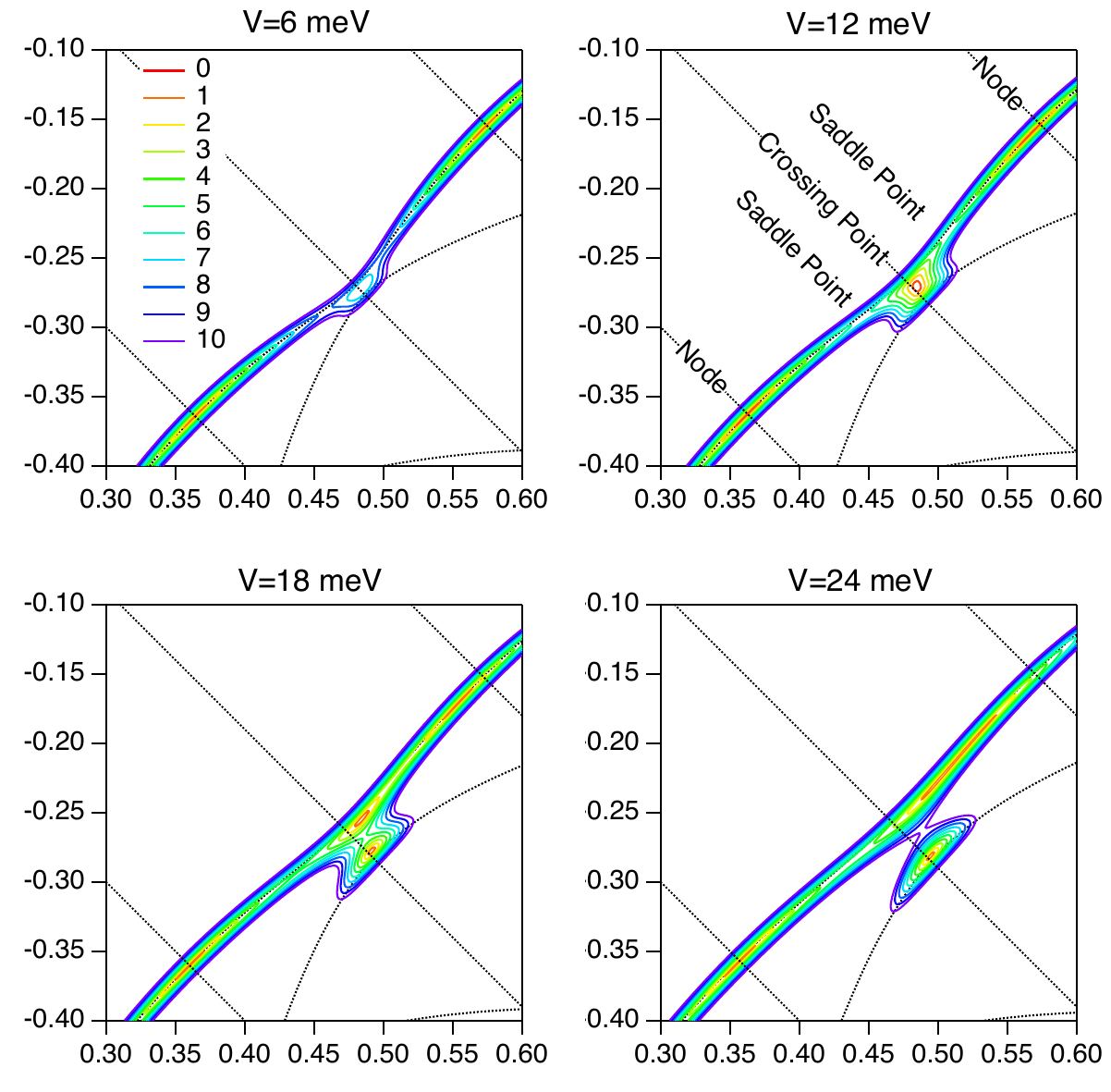}
\caption{Constant energy contours (0 to 10 meV in steps of 1 meV) for the supermodulation case with $\Delta_0$=40 meV and $V$ ranging from 6 to 24 meV.  The dotted curves are the normal state Fermi surfaces, with avoided crossings due to $V$.  The dotted diagonal lines go through the d-wave nodes (lower left and upper right) and through the crossing point (middle).}
\label{fig2}
\end{figure}

To start with, to a good approximation for the cuprates, the d-wave order parameter has the form
\begin{equation}
    \Delta_k = \Delta_0 [\cos(k_xa)-\cos(k_ya)]/2
\end{equation}
At the crossing points, $|\Delta_k| \sim 0.3 \Delta_0$.  For Bi2212, $\Delta_0 \sim 40$meV at optimal doping \cite{Vishik12}.  $V$ has been determined by looking at the avoided crossings in the band structure by photoemission \cite{Valla19,Gao20}.  A value of $V$ of 12.5 meV is observed \cite{Valla19}.  Therefore, at the crossing point, we have that $\Delta_k \sim V$, the significance of which is apparent in Fig.~\ref{fig2}, where the evolution of the low energy structure is shown as $V$ increases from 6 to 12 to 18 to 24 meV.  These plots zoom into a region of the zone that spans from the d-wave node of the original Fermi surface (intersection of the dotted diagonal line with the Fermi surface in the lower left corner of the plots) and the analogous d-wave node of one of the translated surfaces (intersection of the diagonal dotted line with the Fermi surface in the upper right corner of the plots), with the crossing point near the center of the plots (the central diagonal dotted line goes through the crossing point).  For $V=6$ meV, one sees that a minimum in the dispersion has developed at the crossing point which is quadratic in all directions.  Note the crossing point is in between the two Fermi surfaces (shown as dotted curves) as they have been split by $V$.  One can also see the two d-wave nodes in the lower left and upper right corners of this plot.  When $V$ increases to 12 meV, this minimum at the crossing point movers to near zero energy.  Analytically, this is easily understood if one reduces the above secular equation (6 $\times$ 6) to a simpler equation (4 $\times$ 4) that just involves these two bands ($\epsilon_k$ and $\epsilon_{k-Q}$).  Along the diagonal dotted line going through the crossing point (denoted as the crossing line), both $\epsilon$ are equal and both $\Delta$ are equal in magnitude but have opposite sign.  So, the resulting dispersions are $E_k = \pm V \pm \sqrt{\epsilon^2_k+\Delta^2_k}$.  At the crossing point where $\epsilon_k=0$ and $|\Delta_k|=V$, this reduces to 0 and $2V$ for the positive energy Bogoliubov energies.  By expanding the square-root, one finds a quadratic dispersion along the crossing line.  Orthogonal to this line, the dispersion is linear, since the two $\epsilon$ now differ. In this approximation, we note that the results are symmetric relative to the crossing line. In the upper right plot of Fig.~\ref{fig2}, though, one sees the actual minimum has moved slightly off the crossing line.  This is due to the influence of the third band ($\epsilon_{k+Q}$).  In this full 6 $\times$ 6 case, the results instead are symmetric relative to the diagonal dotted line going through the d-wave node of the original Fermi surface (diagonal dotted line in the lower left corner of the plots).  This third-band effect also acts to slightly increase the critical $V$ (denoted as $V_{c1}$) from 12 meV to about 12.5 meV.  As $V$ increases beyond $V_{c1}$, this semi-Dirac node splits into two.  The first one (denoted as $L$) moves approximately along the crossing line to approach the other $V$ split Fermi surface.  The second (denoted as $T$) moves along the first $V$ split Fermi surface and starts to approach the d-wave node of the translated Fermi surface, which itself has started to move towards the crossing line.  At a $V$ just beyond that of the lower right plot of Fig.~\ref{fig2}, these two nodes merge for $V=V_{c2}$, to form a new semi-Dirac node.  This has a quadratic dispersion along the Fermi surface and a linear dispersion orthogonal to it, that is, with a dispersion rotated by $90^\circ$ relative to that of the semi-Dirac node at $V_{c1}$.  For $V$ larger than $V_{c2}$, the semi-Dirac node is lifted.  This then leads to Table 1, where the number of nodes in an $X$ quadrant of the zone is listed as a function of $V$.

\begin{table}
\caption{The number of nodes per $X$ quadrant as $V$ is varied.  Here  $T$ refers to the split node that moves in the transverse direction (along the Fermi surface).}
\begin{ruledtabular}
\begin{tabular}{lll}
$V$  & No. & comments \\
\hline
$V$=0 &  3 & d-wave nodes for $\epsilon_k=0$ and $\epsilon_{k \pm Q}=0$\\
$V < V_{c1}$ & 3 & minima form at the crossing points\\
$V = V_{c1}$& 5 & the minima reach 0 (semi-Dirac nodes)\\
$V_{c1} < V <V_{c2}$ & 7 & the nodes split in orthogonal directions\\
$V = V_{c2}$ & 5 & the $T$ nodes merge with the $\epsilon_{k \pm Q}$ nodes\\
$V > V_{c2}$ & 3 & these semi-Dirac nodes are lifted\\
\end{tabular}
\end{ruledtabular}
\label{table1}
\end{table}

\begin{figure}
\includegraphics[width=\columnwidth]{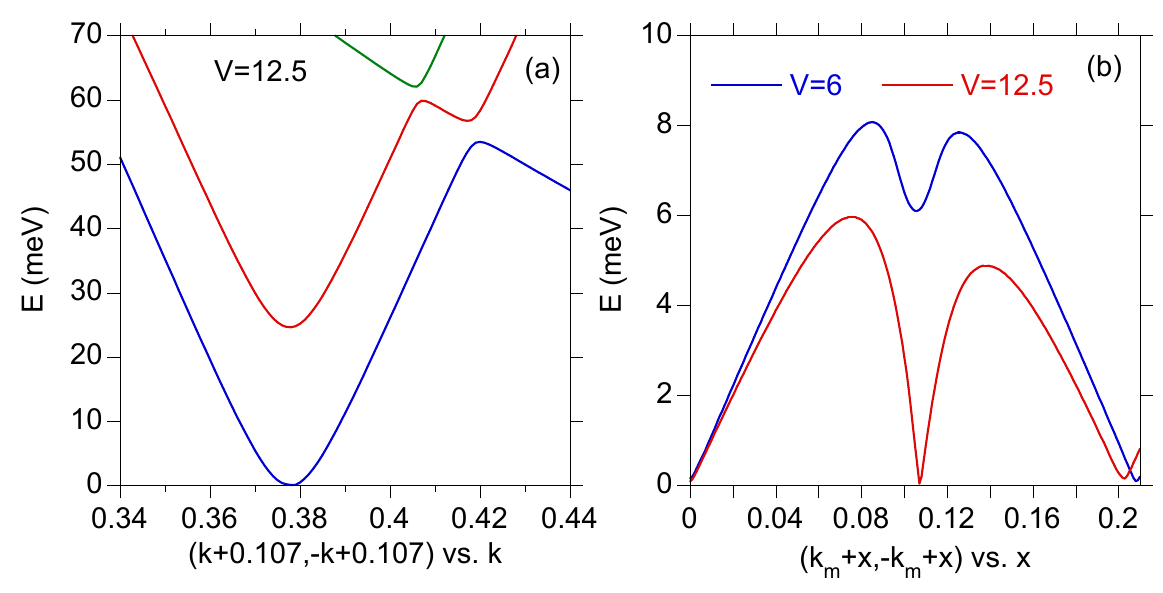}
\caption{(a) Bogoliubov dispersions along a diagonal cut in the zone through the semi-Dirac node for $V=V_{c1}$ (12.5 meV). (b) Minimum value of the Bogoliubov dispersion along parallel diagonal cuts ranging from the nodal direction of $\epsilon_k$ ($x=0$) to that for $\epsilon_{k-Q}$ ($x=0.21$) for $V=6$ and $V$=12.5 meV.  Note the crossing line corresponds to $x=0.105$.}
\label{fig3}
\end{figure}

\begin{figure}
\includegraphics[width=\columnwidth]{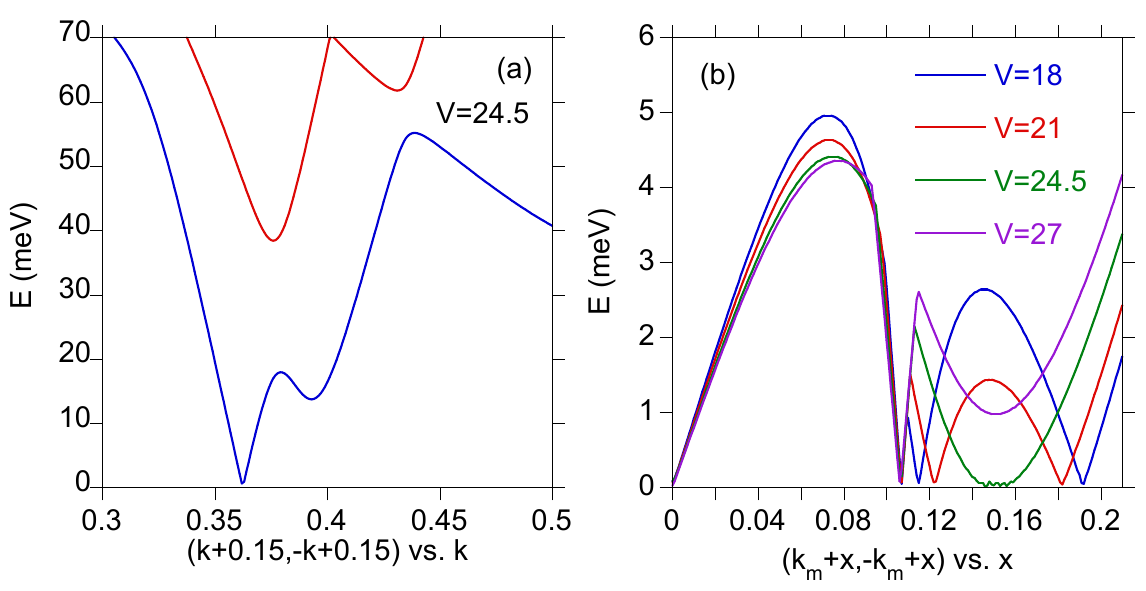}
\caption{(a) Bogoliubov dispersions along a diagonal cut in the zone through the semi-Dirac node for $V=V_{c2}$ (24.5 meV). (b) Minimum value of the Bogoliubov dispersion along parallel diagonal cuts ranging from the nodal direction of $\epsilon_k$ ($x=0$) to that for $\epsilon_{k-Q}$ ($x=0.21$) for various values of $V > V_{c1}$.
Note that one node ($L$) stays near the crossing line ($x=0.105$).  The other ($T$) moves towards the d-wave node (that is at $x=0.21$ for $V=0$).  The two then merge at $V=V_{c2}$.  For larger $V$, this node is lifted.}
\label{fig4}
\end{figure}

These results can be better appreciated from Figs.~\ref{fig3} and \ref{fig4}, where dispersions are shown in two directions for various values of $V$, highlighting the points made above.  In Fig.~\ref{fig3}a, the Bogoliubov dispersions for $V=V_{c1}$ are plotted along a diagonal line in the zone going through the minimum (close to and parallel to the crossing line), showing the quadratic dispersion of the semi-Dirac node along this direction.  In Fig.~\ref{fig3}b, the minimum of the dispersion along each diagonal line of the zone is plotted, showing the linear dispersion of the semi-Dirac node along the Fermi surface, and then the maxima on either side of the semi-Dirac node that mark the saddle points in the dispersion also noted in the upper right plot of Fig.~\ref{fig2}.
In this plot is also shown the minimum for a $V < V_{c1}$ so one can see how the minimum develops at/near the crossing line and reaches zero energy as $V$ approaches $V_{c1}$.  In Fig.~\ref{fig4}b, the same plot is shows but for $V > V_{c1}$ to show how the now split nodes moves with $V$, with the $L$ node being pinned near the crossing line and the $T$ node moving along the Fermi surface where it eventually merges with the $k-Q$ d-wave node for $V \sim 24.5$ meV, forming a quadratic minimum at zero energy.  For larger $V$, one can see the lifting of the merged node.  In Fig.~\ref{fig4}a, the orthogonal direction (along the diagonal line that goes through the node) is shown for $V=V_{c2}$, exhibiting the linear dispersion of the semi-Dirac node along this direction.

\section{Implications}
As mentioned earlier in the text, $V_{c1}$ is approximately realized in bilayer Bi2212.  Since $\Delta_0$ has a strong dependence on doping, then this leads to the possibility of sweeping through $V_{c1}$ by varying the doping.  This could be looked at further via photoemission.  As a caveat, bilayer splitting has been ignored in these calculations.  As this splitting goes as the square of the d-wave order parameter, it is not a big factor given that all the action is near the d-wave nodes.  Still, the square of 0.3 (the ratio of $\Delta_c$ to $\Delta_0$) is 0.09, so it is not entirely negligible (this splitting would quadruple the number of crossing points).  Although we have taken $V$ to be fixed, this could potentially change as well, for instance, if pressure were applied along the $c$-axis.

What about $V_{c2}$?  In single-layer Bi2201, $\Delta_0$ is about half of what it is in Bi2212.  This would imply that Bi2201 is intrinsically near $V_{c2}$. Again doping and/or pressure could be used to sweep through this critical value of $V$.  For both Bi2201 and Bi2212, there could be a signature even in bulk thermodynamics if $V$ is critical in that a semi-Dirac node gives rise to a $\sqrt{E}$ dependence of the density of states around the chemical potential.

We now turn to the relation to the twisted case.  The evolution of the electronic structure with $V$ should be similar to what happens in the trilayer twisted case, but there are some global differences, particularly with respect to the commonly studied twisted bilayer case.  First, the supermodulation acts similarly to a linear charge density wave.  As such, the electronic structure is not chiral in nature.  Second, the novel physics explored in the twisted case due to time reversal symmetry breaking is above and beyond what is presented above.  In these contexts, recent gyrotropic measurements on Bi2201 and Bi2212 indicate that both chiral and inversion symmetry breaking occur \cite{Lim22} as well as mirror symmetry breaking in Bi2212 as observed by second harmonic generation \cite{Jung24}.  Time reversal symmetry breaking has also been observed in Bi2212 from polarized neutron scattering measurements \cite{Didry12,Thro14}.  So, just as in the twisted case, these extra ingredients could give rise to topological superconductivity.

\begin{acknowledgments} 
This work was supported by the Materials Sciences and Engineering Division, Basic Energy Sciences, Office of Science, US Dept.~of Energy.  The author acknowledges discussions with Marcel Franz and Jed Pixley about the relation of these results to those of twisted cuprates.
\end{acknowledgments}

\bibliography{references}

\end{document}